\documentclass[12pt]{article}    

\usepackage{epsfig}

\begin{document} 


\def\ni{\noindent}
\def\beg{\begin{equation}}
\def\neq{\end{equation}}

\def\Xv{{\bf X}}
\def\Yv{{\bf Y}}
\def\rv{{\bf r}}
\def\vv{{\bf v}}
\def\wv{{\bf w}}
\def\av{{\bf a}}
\def\pv{{\bf p}}
\def\qv{{\bf q}}
\def\kv{{\bf k}}
\def\jv{{\bf j}}
\def\Ev{{\bf E}}
\def\Bv{{\bf B}}
\def\Av{{\bf A}}
\def\Rv{{\bf R}}
\def\Pv{{\bf P}}
\def\Kv{{\bf K}}
\def\Qv{{\bf Q}}
\def\sv{{\bf s}}
\def\Sv{{\bf S}}

\def\xu{\hat{\bf x}}
\def\yu{\hat{\bf y}}
\def\zu{\hat{\bf z}}
\def\uu{\hat{\bf u}}
\def\ie{{\it i.e.}}
\def\fv{{\bf f}}
\def\ru{\hat{\bf r}}
\def\ku{\hat{\bf k}}
\def\nv{\hat{\bf n}}
\def\bv{{\bf b}}

\def\polt{\Sigma}
\def\polv{\vec{\cal S}}
\def\pol{{\cal S}}
\def\poltr{{\cal T}}
\def\SL{{\rm S}_{\rm L}}
\def\ST{{\bf S}_{\rm T}}
\def\X{{\rm X}}
\def\T{_{\rm T}}
\def\L{_{\rm L}}
\def\A{_{\rm A}}
\def\B{_{\rm B}}
\def\uA{^{(A)}}
\def\uB{^{(B)}}
\def\q{^{(q)}}

\def\plusdemi{|+{\scriptstyle{1\over2}}\rangle}
\def\moinsdemi{|-{\scriptstyle{1\over2}}\rangle}      
\def\plusun{|+1\rangle}
\def\moinsun{|-1\rangle}
\def\plus{|+\rangle}
\def\moins{|-\rangle}      
\def\Xrond{{\rm X}\raise1.8ex\hbox{\kern-0.6em\hbox{$\circ$}}\ }
\def\krond{{\bf k}\raise1.8ex\hbox{\kern-0.6em\hbox{$\circ$}}\ }


     
\vspace{2cm}


\noindent
{\large The Transverse Spin}

\vspace{5mm}
\noindent
{\large X. ARTRU}

\vspace{5mm}
\noindent
{\em  Institut de Physique Nucl\'eaire de Lyon, \\
IN2P3-CNRS and Universit\'e Claude Bernard, \\
F-69622 Villeurbanne, France}


\vspace{1.5cm}

\noindent
{\bf Contents :}

\smallskip
\noindent
1. Pre-history

\noindent
2. Transversity versus helicity

\noindent
3. The massless limit. "Cardan" and "see-saw" transformations

\noindent
4. Transversity distribution $\delta q(x)$. The diquark spectator model

\noindent
5. Soffer inequality

\noindent
6. Tensor charge sum rule

\noindent
7. t-channel analysis

\noindent
8. Selection rules for $\delta q(x)$ measurements

\noindent
9. Evolution with $Q^2$

\noindent
10. Quark polarimetry. The sheared-jet (Collins) effect

\noindent
11. Single-spin asymmetries in inclusive experiments

\noindent
12. Quark distribution dependent on both spin and $\kv\T$ 

\noindent
13. First evidence of quark transversity


\bigskip

\noindent
{\bf 1. Prehistory of transverse spin}

\noindent
Some natural or artificial moving objects,
like the neutrino and the propellers of air-planes or boats,
rotate about axes parallel to their velocities~: they posses 
longitudinal spin or {\it helicity}.
Others, like petanque bowls, charriot wheels and paddle wheels
rotate about axes perpendicular to their velocities~:
they posses transverse spin or {\it transversity}.
Comparing these two classes of objects,
an "alphysicist" (the analogue of an alchimist)
would remark that ``transversity is a virtue of slow or heavy objects, 
helicity is a virtue of fast or light objects''. 
Physicists are not alphysicists, however they have been deluded by an 
apparent quenching of transversity in high-energy physics~:

\begin{itemize}
\item the spin polarization of a relativistic particle can be described
by the 4-vector $\pol^\mu$ (see \S 2).
If we let the momentum go to infinity, 
the ratio between the transverse to the longitudinal components of $\pol$ 
go to zero.

\item{} in {\it fully inclusive} deep inelastic lepton scattering (DILS), 
the transverse spin asymmetries
are negligible compared to the usually measured longitudinal spin asymmetry.

\end{itemize}

\noindent
So, during the 1970's and 1980's there was a
prejudice that transverse spin is small or irrelevant for ultrarelativistic
($m \ll E$) particles, or at least in hard reactions 
($m^2 \ll Q^2$, where $Q$ is the deep inelastic momentum scale).
This prejudice persisted in spite of large transverse polarizations 
in $e^+$ and $e^-$ storage rings produced by the Sokolov-Ternov effect 
(see \S7.3 of Ref.\cite{CHJR}).
Also experimental single-spin asymmetries
with transversely polarized initial protons (\cite{adams} and refs. therein)
or final Lambda's \cite{Felix}
were repeatedly reported, but they were believed to be the "tail" 
of a low-energy phenomenon.

During that period, the physics of transverse spin progressed through
isolated works.
Studying the Drell-Yan reaction with polarized protons 
\begin{equation}
\uparrow\! A + \uparrow\! B \to l^+ l^- + {\rm anything}
\,,
\label{D-Y}
\end{equation}
Ralston and Soper \cite{ralston1} 
introduced the transversity distribution $\delta q(x)$
and predicted a double spin asymmetry for transversely polarized baryons~:
the polarized differential cross section writes
$$
d\sigma^{(D-Y)}_{\rm pol.}
= dx_a\;dx_b\; \sum_a
d\hat \sigma_{\rm unpol.}^{(a+b\to l^+ l^-)} \times\  \big[a(x_a)\;b(x_b)
-\SL\uA \; \SL\uB \;\Delta a(x_a)\;\Delta b(x_b) 
$$
\begin{equation}
-|\ST\uA|\;|\ST\uB|\;\delta a(x_a)\;\delta b(x_b)\;
\hat A_{NN} (\hat \theta)\;\cos (2\phi-\phi_A-\phi_B) \big]
\,.
\label{Ral-Sop}
\end{equation}
The notations are as follows~:
$a(x_a)$ is the unpolarized distribution of quark species $a$
($a = u, d, s, \bar u, \bar d, \bar s$) in proton $A$,
where $x_a=p_L^{(a)}/p^{(A)}$ is the scaled longitudinal momentum
(similarly with $b$ in proton $B$).
The sum is made over the parton species, with $a=\bar b$.
$\SL\uA$ and $\ST\uA$ are the longitudinal 
and transverse polarizations of $A$ 
($\SL\equiv 2 \times$ average helicity, $\ST\equiv 2 \times$ average
transverse spin).
$\Delta a(x)$ and $\delta a(x)$ are the helicity and 
transversity distributions~:
\begin{equation}
\Delta a(x) = a^+(x) - a^-(x)
\,,\quad
\delta a(x) = \ \uparrow\! a(x) - \downarrow\! a(x) \,,
\label{poldistrib}
\end{equation}
$a^\pm(x)$ being the distributions 
of quark with helicities $\pm{1\over2}$ in 
a baryon of helicity $+{1\over2}$,
while $\uparrow\! a(x)$ and $\downarrow\! a(x)$ 
are analogous quantities for transverse polarization.
$d\hat\sigma$ is the cross section of the subprocesss and 
\begin{equation}
\hat A_{NN}(\hat\theta)\ =\
- 2\,\hat t \hat u / (\hat t^2 + \hat u ^2)\ =\   
-\sin^2\hat\theta/ (1+\cos^2\hat\theta)
\label{Ann}
\end{equation}
the associated double spin asymmetry,
for quarks polarized normal to the scattering plane.
$\phi$ is the azimuth of the $l^+ l^-$ relative momentum
and $\phi_A$ and $\phi_B$ the azimuths of $\ST\uA$ and $\ST\uB$.
Formula (\ref{Ral-Sop}) teaches us that~:

\begin{itemize}
\item{} {\it transverse spin asymmetry is not suppressed by a power of $Q$}
(here $Q = l^+l^-$ mass).
\item{} {\it it disappears if one integrates over $\phi$} 
(Hikasa theorem\cite{Hikasa}).
\end{itemize}

\noindent
Other authors \cite{CHJR,qq-scat,Baldra} 
considered double transverse spin asymmetries 
in inclusive reactions. However the prejudice of a small 
$\delta a(x)$ or its confusion with $g\T$ (see end of \S2) were still alive.

Now the situation has completely changed and transverse quark polarization
is a well-considered field. Several international workshops have been 
dedicated to it \cite{RIKEN,Zeuthen}. 

The present paper is more an introduction than a review. Many important 
works are not mentionned and we apologize for missing or incomplete references.
We tried to minimize the number of notations and
equations, but added some personal points of view, e.g., about
the $m\to0$ limit and about the relevance of transversity to electric 
dipole moments.
An excellent and well-documented review is given in Ref.\cite{barone}.

\newpage
\noindent
{\bf 2. Transversity versus helicity}

\noindent
Consider first a non-relativistic bound state, e.g., an atom.
According to the Wigner-Eckart theorem the average electron spin
is colinear to the atom spin~:
$$
\langle \sv_e \rangle = C \ \langle \sv_A \rangle
\,,
$$
independently of the relative orientation of the atom spin
$\sv_A$  and momentum $\pv_A$. 
The electronic contributions to the atom helicity and transversity are 
therefore equal~:
\begin{equation}
\Delta n_e = \delta n_e = C \, n_e
\,.
\label{Wigner-E}
\end{equation}
If spin is decoupled from orbital motion, we can generalize (\ref{Wigner-E})
to the unpolarized and polarized electron densities 
$n_e(\kv)$, $\Delta n_e(\kv)$ and $\delta n_e(\kv)$
in the internal momentum space.
If however there is a spin-orbit coupling, we cannot
apply the Wigner-Eckart theorem at fixed $\kv$ and we have generally 
\begin{equation}
\Delta n_e(\kv) \ne \delta n_e(\kv)
\,.
\label{spin-orbit}
\end{equation}

\noindent
Consider now the relativistic case. The polarization vector
$\Sv = 2 \langle \sv \rangle$ of a spin ${1\over2}$ particle can be 
generalized in two ways~:

\smallskip
\noindent
{\it A) as a pseudo-vector}. In the rest frame of the particle, we promote 
$\Sv$ to the 4-vector $\pol^\mu = (0,\Sv)$.
In another frame, $\pol^\mu$ has a non-zero time component 
and satisfies
$$
\pol\cdot p = 0
\,,\quad
- \; \pol\cdot\pol \, \equiv \, \polv\cdot\polv-(\pol^0)^2 
\, = \, \Sv^2_{rest} \, \le 1 
\,.
$$
The Pauli-Lubanski vector \cite{Martin} is $m\pol^\mu$.
For a {\it free} Dirac spinor at rest, we have 
\begin{equation}
\pol^x = {\psi^\dagger \, \sigma^x \, \psi \over \psi^\dagger \psi}
= {\bar\psi \, \gamma^x \, \gamma^5 \, \psi \over \bar\psi \psi}
\,,\ etc.,
\quad
\pol^0 = {\bar\psi \, \gamma^0 \, \gamma^5 \, \psi \over \bar\psi \psi}= 0
\,,
\label{Vrest}
\end{equation}
where $\sigma^x\equiv\sigma^{yz}$ and 
$\bar\psi\equiv\psi^\dagger\gamma^0$. In the last two denominators
we have replaced $\psi^\dagger$ by $\bar\psi$ since they coincide in the rest
frame. The generalization to any frame is then
\begin{equation}
\pol^\mu 
= {\bar\psi \, \gamma^\mu \, \gamma^5 \, \psi \over \bar\psi \psi} \,.
\label{V}
\end{equation}

\smallskip
\noindent
{\it B) as an antisymmetric tensor}. 
Since $\bar\psi=\psi^\dagger$ at rest, the first numerator of (\ref{Vrest}) 
can be written $\bar\psi \, \sigma^{yz} \, \psi$ as well. 
Thus we define
\begin{equation}
\polt^{\mu\nu} = {\bar\psi \, \sigma^{\mu\nu} \, \psi \over \bar\psi \psi}
\,.
\label{T}
\end{equation}
For a particle at rest,
$\polt^{yz} = \pol^x$, etc., $\polt^{0i} = 0$.
In an arbitrary frame, 
\begin{equation}
\pol^\mu = {1\over2m} \, \varepsilon^{\mu\nu\rho\lambda}
\, p_\nu \, \polt_{\rho\lambda}
\,,\qquad
\polt^{\mu\nu} = {1\over m} \, \varepsilon^{\mu\nu\rho\lambda}
\, p_\rho \, \pol_\lambda
\qquad 
(\varepsilon^{0123} = -1).
\label{corres}
\end{equation}
Thus, for an on-mass-shell ({\it i.e.}, free) particle,
(\ref{V}) and (\ref{T}) are in one-to-one correspondence. 
This remains approximately true for a weakly bound particle,
because $\bar\psi\simeq\psi^\dagger$ in the bound-state frame.
However, the correspondence breaks down for relativistically
bound particles like quarks in a hadron. 
These quarks are significantly off-mass shell, their Dirac spinors 
have 4 independent components instead of two and the 
independent bilinear observables are 16 instead of 4.
Thus, there is an ambiguity about which matrix,
$\gamma^\mu\gamma^5$ or $\sigma^{\mu\nu}$, 
describes the quark polarization in a hadron.

The "leading twist" prescription is a mixture of the two~:
quark helicity will be defined in terms of 
$\bar\psi\gamma^z\gamma^5\psi$
whereas quark transversity will be defined in terms of
$\bar\psi\sigma^{yz}\psi$ and $\bar\psi\sigma^{zx}\psi$
(from now on, we assume that the hadron is moving in the $+\zu$ direction). 
To understand it, let us look at the free particle case again.
After a Lorentz boost along $\zu$, we have 
\begin{equation}
\pol^i = \pol_{rest}^i
\,,\quad \pol^z = \gamma \ \pol_{rest}^z
\,,
\label{boost1}
\end{equation}
\begin{equation}
\polt^{iz} = \gamma \ \polt_{rest}^{iz}
\,,\quad \polt^{xy} = \polt_{rest}^{xy}
\,,
\label{boost2}
\end{equation}
where $i=x$ or $y$.
Thus, $\pol^z$, $\polt^{yz}$ and $\polt^{zx}$  are amplified
by the Lorentz factor $\gamma=E/m$, whereas 
$\pol^x$, $\pol^y$ and $\polt^{xy}$ are not.
A similar phenomenon occurs with interacting quarks~:
considering the bilinear operators
$\bar\psi(X') \; \Gamma \; \psi(X)$,
amplification occurs for
$\Gamma = \gamma^z\gamma^5$, $\sigma^{yz}$ or $\sigma^{zx}$,
not for $\Gamma = \gamma^x\gamma^5$, $\gamma^y\gamma^5$ or $\sigma^{xy}$
(here $\psi(X)$ is the quark {\it field} operator).
At leading order in $1/Q$, polarized experiments are only sensitive
to the amplified bilinear operators.
This is why $\Gamma = \gamma^z\gamma^5$ is used to define 
the quark helicity distribution $\Delta q(x)$
whereas $\Gamma = \sigma^{yz}$ or $\sigma^{zx}$ 
is used to define the quark transversity distribution $\delta q(x)$ 
(see Refs.\cite{ralston2, -sity}).

The name {\it tranversity} was advocated by Jaffe and Ji \cite{-sity}, 
instead of {\it transverse spin}~; 
indeed, the spin density along $\xu$ is 
$\psi^\dagger\sigma^{yz}\psi \equiv \bar\psi\gamma^x\gamma^5\psi$ 
and not $\bar\psi\sigma^{yz}\psi$. One $\gamma^0$ matrix makes the difference.
The "transverse spin" distribution in the strict sense is 
$g\T(x)=g_1(x)+g_2(x)$, built with $\Gamma = \gamma^\perp\gamma^5$
\cite{jajig2}.
However, when only leading twist effects are considered,
"transverse spin" is often used - improperly - in place of "transversity".

Naively, we would have
$\delta q(x) = g\T(x) = \Delta q(x) = h\L(x)$,
the latter being built with $\Gamma = \sigma^{xy}$.
However,

\noindent
1) the generalization of (\ref{corres})
to $\bar\psi(X')\gamma^\mu\gamma^5\psi(X)$
and $\bar\psi(X')\sigma^{\mu\nu}\psi(X)$
is not valid for strongly bound quarks. Thus,
$\delta q(x) \ne g\T(x)$ and $\Delta q(x) \ne h\L(x)$.

\noindent
2) in the nonrelativistic SU(6) model, quarks are in S-waves. 
In a relativistic model, the Dirac wave function
has also a so-called "small" component, which is a P-wave 
and possesses a spin-orbit coupling. Therefore, once the  
lightcone momentum $k^0+k^z = x \; (p^0 + p^z)$ is fixed,
we cannot apply the Wigner-Eckart theorem. 
By analogy with (\ref{spin-orbit}), we have
$\Delta q(x) \ne g\T(x)$ and $\delta q(x) \ne h\L(x)$.
To conclude, 
$$
\delta q(x) \ne g\T(x) \ne \Delta q(x) \ne h\L(x)
\,.
$$

\medskip
\noindent
{\bf 3. The massless limit}

\noindent
Although the quark mass is nonzero, the $m\to0$ limit
is worthwhile to study. The basic fermions of the standard model 
are indeed massless and obey chiral symmetry. 
This r\'egime is overthrown by the Higgs mechanism, 
but it can be approximately restored in deep inelastic sub-processes, 
where masses can be neglected compared to momenta.

For a massless particle, we cannot start from the rest frame to define 
the polarization, as we did before. Nevertheless, the
helicity states $\plusdemi$ and $\moinsdemi$ are well-defined,
and we can construct the transversity states as linear
combinations of them. For instance,  
\begin{equation}
|\pm\xu\rangle = 2^{-1/2} \, \left( \,
\plusdemi \pm \moinsdemi \, \right)
\label{mixt}
\end{equation}
{\it in principle} defines a state polarized toward $\pm\xu$
(the restriction {\it "in principle"} will be explained later).
This has to be compared to the photon case, where
$$
|\xu\rangle = 2^{-1/2} \, ( \,
\plusun + \moinsun \, )
\,,\quad
|\yu\rangle = 2^{-1/2} \, ( \,
\plusun - \moinsun \, )
$$
are the states of {\it linear} polarization along $\xu$ and $\yu$.
More generally, for a fermion, 
\begin{equation}
|\nv\rangle = 2^{-1/2} \, \left( \,
\plusdemi + e^{i\phi} \moinsdemi \, \right) \,,
\label{mixt/phi}
\end{equation}
where $\phi$ is the azimuth of the transverse unit vector $\nv$,
whereas for the photon 
$$
|\nv\rangle = 2^{-1/2} \, ( \,
\plusun + e^{2i\phi} \moinsun \, )
\,.
$$

If there were no Higgs mechanism mixing the states
$\plusdemi$ and $\moinsdemi$, 
their relative phase could be changed arbitrarily (chiral transformation)
and Eq.(\ref{mixt/phi}) would not define the azimuth of the transversity vector
unambiguously. This is the reason of our restriction 
{\it "in principle"} above. An analogous phenomenon occurs
in electrodynamics in the absence of charges and currents~:
the Maxwell equations would be invariant under chiral rotations
mixing $\Ev$ and $\Bv$ and there would be no way 
to define the azimuth of the linear polarization.

For $m\ne0$ and $\pv \parallel \zu$,
$$
\pol = \left( \SL {p^z \over m}, \; \ST, \;   \SL {p^0 \over m} \right)
\,,
$$
where $\ST$ is the transversity vector and $\SL$ the longitudinal polarization,
which obey 
\begin{equation}
\SL^2 + \ST^2 \le 1  \,,\quad \pv \cdot \ST = 0
\,,
\end{equation}
the inequality being saturated for a pure state.

In the limit $m\to0$ at fixed $\pv$, as in (\ref{boost1}),
the transverse part of $\pol$ does not change, 
but the time+longitudinal part becomes infinite as
$m^{-1} \, \SL \, p^\mu$.
Then $\SL$ can be defined as $\lim_{m\to 0} (m \, \pol^0 / p^0)$ 
and becomes invariant under all Lorentz transformations.
As for $\ST$, it can be promoted to a 4-vector $\poltr^\mu$, 
obtained from $\pol$ by subtracting the infinite time+longitudinal part~:
$$
\poltr^\mu = \lim_{m\to0} \, \pol^\mu - m^{-1} \, \SL \, p^\mu \,.
$$
This limit is of the type $\infty - \infty$, therefore undetermined. 
Due to the division by $m$, an infinitesimal error on $\SL$ 
can produce a finite error on $\poltr^\mu$ of the form $C\, p^\mu$. 
We give up determining $\poltr$ completely and define it only up 
to a "gauge" transformation~:
\begin{equation}
\poltr^\mu \rightarrow \poltr^\mu + C \, p^\mu
\,.
\label{gauge}
\end{equation}
$\poltr = (0, \ST)$ is just a particular "gauge". 
To calculate the effect of a non-colinear Lorentz transformation on $\ST$,
we first transform $\poltr$
into $\poltr'$, then eliminate the time and longitudinal components
of $\poltr'$ using (\ref{gauge}) with $C=-\poltr'^0/p'^0$
(the same method works for the polarization vector of a photon).
Gauge-like invariance can also be used to test polarization
formulas at $m\to0$.

The covariant density matrix $\hat \rho = u \, \bar u$
(with the normalization $u^\dagger \, u = 2E$)
has a finite limit for $m\to0$ \cite{landau}~:
$$
\hat\rho = {1 - \gamma \cdot \pol \ \gamma^5 \over 2} \, (\gamma\cdot p + m)
\quad \to \quad
{1 - \gamma \cdot \poltr \, \gamma^5 + \SL \, \gamma^5 \over 2} 
\, \gamma\cdot p
$$
which is invariant under (\ref{gauge}). This formula involves 
helicity and transversity quite differently.

\medskip
\noindent
{\it ``Cardan'' and ``see-saw'' transformations} (p.323 of \cite{RIKEN})

For $m=0$, QCD and electroweak interactions conserve fermion helicities~:
an initial fermion $f$ and a final fermion $f'$ connected by an internal 
fermion line have equal helicities~; 
a fermion $f_1$ and an antifermion $\bar f_2$
connected together have opposite helicities.
Due to this rule, transversely polarized cross sections are invariant when 
all the transversity vectors are simultaneously rotated 
by an angle $\Delta\phi$ about the respective particle momenta. 
The reader can check it using (\ref{mixt/phi}).
We call this a "cardan transformation", by analogy
with rotations transmitted by a mechanical cardan
(for field theorists, it is nothing but a chiral transformation).
Separate cardan transformations can also be applied 
to disconnected subsets of external fermions.
The dependence of (\ref{Ral-Sop}) on $\phi_A+\phi_B$ only, not on $\phi_A$
and $\phi_B$ separately, is a consequence of this invariance.

{\it Oblique} polarizations states can be written as 
\begin{equation}
|\SL, \ST \rangle = 
2^{-{1\over2}} \, |\ST|^{1\over2} \,
\left( e^{\eta - i\phi \over 2}   \plusdemi 
+ e^{i\phi -\eta \over 2} \moinsdemi \, \right)
\,,
\end{equation}
with $\SL \equiv \tanh \eta$ and $|\ST|=1/\cosh\eta$.
Cardan rotations $\phi_i \to \phi_i + \Delta \phi $ can be 
completed by {\it see-saw} transformations
\begin{equation}
\eta_i \to \eta_i + \Delta \eta 
\,,\quad  \eta'_j \to \eta'_j - \Delta \eta \,,
\end{equation}
where $f_i$ are initial fermions, $f'_j$ are final ones.
What is invariant under these transformations is the reduced cross section
$$
\tilde\sigma \equiv 
\left( \prod_i |\ST^{(i)}| \, \prod_j |\ST'^{(j)}| \right)^{-1} \ \sigma
\,.
$$

\begin{figure}
\centering\epsfig{file=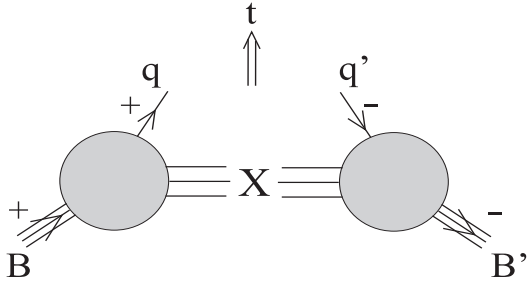,width=7cm,height=4cm}
\caption{Unitarity diagram representing $\delta q(x)$,
according to (\ref{interf}).}
\end{figure}

\newpage
\noindent
{\bf 4. The transversity distribution}

\noindent
In the infinite momentum frame, the momentum density of quarks 
of polarization $\Sv_q$ in a baryon of polarization $\Sv_b$ is
\begin{equation}
{dN\over dx \, d^2\kv\T} = 
\sum_\X \left| \,
\langle \kv,\Sv_q \, ; \, \X \, | \,
\pv, \Sv_B \rangle \, \right|^2
\,,\label{distrib}
\end{equation}
where $|\X\rangle = |\kv_1 \Sv_1, \kv_2 \Sv_2 ...\rangle$ is a multi-parton
spectator state. Normalization factors as well as a factor
$\delta^{(3)}(\kv + \pv_\X - \pv)$ are omitted.
We introduce the short-hand notation
\begin{equation}
\langle \kv,\Sv_q \, ; \, \X | \pv, \Sv_B \rangle
\equiv \langle \kv, \Sv_q | V(\X) | \pv, \Sv_B \rangle
\equiv \X_{\Sv_q\Sv_B}
\,.
\label{matriX}
\end{equation}
The first equality defines a generalized vertex operator
${ V}(\X)$ connecting quark and baryon states.
In the last expression we have replaced ${ V}(\X)$ by $\X$
to shorten the writing and omitted the arguments $\pv$ and $\kv$,
since $\pv$ is fixed and $\kv \equiv \pv - \pv_\X$.
The matrix $\X$ in the spin sub-space 
is linked to the state $|\X\rangle$.
In the transversity basis (\ref{mixt}), we rewrite Eq.(\ref{poldistrib}) as
$$
\delta q(x,\kv\T) = \sum_\X 
\left|\X_{\uparrow\uparrow}\right|^2
- \left|\X_{\downarrow\uparrow}\right|^2
\,.
$$
In the helicity basis, it becomes
\begin{equation}
\delta q(\X,\kv\T) = \sum_\X 
{\rm Re} \, \{\X_{++} \, \X^*_{--} +  \X_{++} \, \X^*_{-+}
    + \X_{+-} \, \X^*_{--} +  \X_{-+} \, \X^*_{+-} \}
\,.\label{interf}
\end{equation}
Due to the invariance upon rotations about $\zu$, 
only the first term survives the integration over $\X$ 
and the azimuth of $\kv\T$~:
Let $R$ be such a rotation, which transforms 
the vector $\krond$ of azimuth zero to $\kv$ of azimuth $\phi$~:
$$
R \; | \krond,\lambda_q \, ; \, \Xrond \rangle
= e^{-i\phi\lambda_q} \;
| \kv,\lambda_q \, ; \, \X \rangle 
\,,\quad
R \, |\pv, \lambda_B \rangle = 
e^{-i\phi\lambda_B} |\pv, \lambda_B \rangle
\,,
$$
where $\kv = R \krond$ and $| \X \rangle$ = $R \, | \Xrond \rangle$.
Then, from (\ref{matriX}),
\begin{equation}
\X_{\lambda_q\lambda_B} \, \X^*_{\lambda'_q\lambda'_B}
= e^{i\phi(\lambda_B - \lambda_q - \lambda_B' + \lambda_q')}
\, \Xrond_{\lambda_q\lambda_B} \, \Xrond^*_{\lambda'_q\lambda'_B}
\,.
\label{binome}
\end{equation}
Putting (\ref{binome}) into (\ref{interf}) 
and replacing the summation on $\X$ by a summation on $\Xrond$,  
one can verify that the last three terms disappear
after integration over $\phi$. Thus
\begin{equation}
\delta q(x) = \int d^2\kv\T \sum_\X 
{\rm Re} \, \{\X_{++} \, \X^*_{--} \}
\,.\label{flip}
\end{equation}
The right-hand side is represented in Fig.1.  
It illustrates that transverse spin asymmetries result from the interference 
between different helicity amplitudes. This comes from 
the quantum superposition (\ref{mixt/phi}).

\medskip
\noindent
{\it Covariant (quark} + {\it scalar diquark) model} \cite{A+M,qq-axial}

The simplest spectator system is a spin zero, positive parity $(qq)$ state,
like for instance the $(ud)$ pair of the $\Lambda$ baryon in the $SU(6)$ model.
Then (\ref{matriX}) is proportional to
$$
\X_{\Sv_q\Sv_B} \propto {g(k^2)\over k^2-m^2_q} \,
\bar u(\kv ,\Sv_q)\;u(\pv,\Sv_B) \,,
$$
where
$k^2=xm_B^2-(k_T^2+xm_{qq}^2)/(1-x)$ is the 4-momentum squared
of the active quark, assumed to be of-mass-shell 
(the spectator is on mass-shell) and $g(k^2)$ is
a baryon-quark-diquark vertex function. In the helicity basis,
\begin{equation}
( \ \X_{\lambda_q\lambda_B} \ ) = C \, {g(k^2)\over k^2-m^2_q} 
\, \pmatrix{
xm_B+m_q & k_x-ik_y   \cr
k_x+ik_y   & xm_B+m_q \cr}
\,,
\label{qq-vertex}
\end{equation}
with $C^2 = x \times [16 \pi^3 (1-x)]^{-1}$. One obtains 
$$
\delta q(x) = q^+(x) = 
C^2 \, (xm_B+m_q )^2 \,
\int d^2 \kv\T \, \left\vert{g(k^2)\over k^2-m^2_q}\right\vert^2 \,,
$$
$$
q^-(x) = C^2 \, 
\int d^2 \kv\T \ \kv\T^2 \, \left\vert{g(k^2)\over k^2-m^2_q}\right\vert^2
\,,
$$
where $q^\pm(x) = {1\over2} [q(x) \pm \Delta q(x)]$ (cf. Eq.\ref{poldistrib}). 
We see that $\delta q(x) \ne \Delta q(x)$.
The equality $\delta q(x) = q^+(x)$ is particular to this model 
and comes from $\X_{++} = \X_{--}$.
This model has been improved \cite{qq-axial} by adding $1^+$ spectator diquarks
which are present in the $SU(6)$ wave functions of the nucleon.
However, it has no sea, no gluon and no confinement. 

\bigskip
\noindent
{\bf 5. The Soffer inequality}

\noindent
Using the general inequality $2 |{\rm Re} \, (ab^*)| \le |a|^2 + |b|^2|$
in (\ref{flip}), we derive
$$
|\delta q(x)| \, \le \, {1\over2} \int d^2\kv\T \sum_\X 
\left( |\X_{++}|^2 + |\X_{--}|^2 \right)
\,.
$$
Parity conservation imposes 
$\sum_\X |\X_{++}|^2 = \sum_\X |\X_{--}|^2 = q^+(x, \kv\T)$.  
Integrating over $\kv\T$, one gets the {\it Soffer inequality} \cite{soffer}
\begin{equation}
|\delta q(x)| \le q^+(x)
\,.
\end{equation}
This inequality is saturated in the quark + scalar diquark model.
In fact it is saturated if (but not only if) there is only one spectator
state $|\X\rangle$. In that case the spectator cannot keep quantum information 
to itself, and the correlation between $\sv_q$ and $\sv_B$ is maximum.


\medskip
\noindent
{\bf 6. The tensor charge sum rule} \cite{ralston2, -sity}

\noindent
\def\charge{{\cal Q}}
The transversity distribution obeys the following sum rule
\begin{equation}
\int dx \, [\delta q(x) - \delta\bar q(x)] = {
\langle B | :\bar\psi(0) \, \sigma^{\mu\nu} \, \psi(0): | B \rangle
\over \bar u_B \, \sigma^{\mu\nu} \, u_B} 
\equiv \delta \charge
\,,
\label{sum}
\end{equation}
where $| B \rangle$ is the baryon plane wave
$u_B \, e^{i\pv \cdot \rv - iEt}$.
$\delta \charge$ is called the {\it tensor charge} of the baryon.
This sum rule is analogous to the vector- and axial- charges sum rules
for $q(x) - \bar q(x)$ and $\Delta q(x) + \Delta\bar q(x)$,
where $\sigma^{\mu\nu}$ is replaced by
$\gamma^\mu$ and $\gamma^\mu \, \gamma^5$ respectively.
Note that quark and antiquark contribute to (\ref{sum}) in opposite ways, 
due to the odd charge conjugation of the $\sigma^{\mu\nu}$ matrix.

The same matrix appears in the coupling of the electromagnetic
field $F^{\mu\nu}$ to an anomalous magnetic moment 
(a.m.m.) or to an electric dipole moment (e.d.m.). 
Therefore, if the quark a.m.m and e.d.m. were
tunable parameters, one would naively expect (\cite{camb} 
and p.324 of \cite{RIKEN}; see also \cite{Chen})
\begin{equation}
  {\partial \hbox{(nucleon a.m.m.)} \over \partial \hbox{(quark a.m.m.)}}
= {\partial \hbox{(nucleon e.d.m.)} \over \partial \hbox{(quark e.d.m.)}}
= \delta \charge
\,.
\label{edm}
\end{equation}

\begin{figure}
\centering\epsfig{file=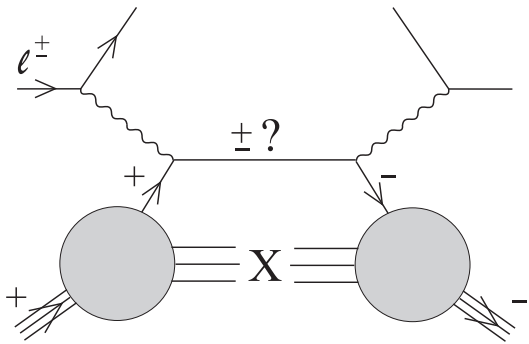,width=7cm,height=4.5cm}
\caption{Unitarity diagram showing the impossibility
to measure $\delta q(x)$ in deep inelastic lepton scattering.}
\end{figure}

\newpage
\noindent
{\bf 7. t-channel analysis} 

\noindent
The complete spin dependence of the quark distribution (\ref{distrib}) 
is contained in the {\it correlation matrix}
\begin{equation}
C_{\lambda_q\lambda_A\lambda'_q\lambda'_A} =
\sum_X \X_{\lambda_q\lambda_A} \, \X^*_{\lambda'_q\lambda'_A} \,.
\label{correl}
\end{equation}
$C$ is the discontinuity in $(p-k)^2 = m_\X^2$
of the quark-baryon forward scattering amplitude
(neglecting quark confinement), 
as illustrated by the unitarity diagram of Fig.1.
As learned while deriving (\ref{flip}), an asymmetry
upon reversal of the quark transversity 
is an interference between amplitudes with $\lambda_q = +{1\over2}$ and 
$\lambda'_q = -{1\over2}$. 

In the {\it t-channel} reaction $B \bar B \to q \bar q$,
indicated by vertical arrow in Fig.1, $q'$ becomes $\bar q$ 
with $\lambda_{\bar q} = - \lambda'_{q}$ 
Therefore the total helicity of the $(q\bar q)$ t-channel state is 
$$
\lambda_q + \lambda_{\bar q} = \lambda_q - \lambda'_{q} = \pm 1
$$
(similarly for $B'$ and $\bar B$, since the baryon is also
transversely polarized).
$\lambda_q - \lambda'_{q}$ is the {\it helicity flip}
in the physical channel.
We can say that transversity asymmetries are associated to
the ``helicity-triplet'' t-channel states 
\begin{equation}
|1, + 1\rangle = \plus \otimes \plus
\,,\quad
|1, - 1\rangle = \moins \otimes \moins
\label{triplet-T}
\end{equation}
(here $|\pm\rangle \equiv |\pm {\scriptstyle{1\over2}}\rangle$).
By comparison, helicity asymmetries correspond
to the ``helicity-triplet'' state 
\begin{equation}
|1, 0\rangle = 2^{-{1\over2}} \,
(\plus \otimes \moins - \moins\otimes \plus)
\label{triplet-L}
\end{equation}
and the unpolarized cross section to the ``helicity-singlet'' state
\begin{equation}
|0, 0\rangle = 2^{-{1\over2}} \, 
(\plus \otimes \moins + \moins\otimes \plus)
\,.
\label{singlet}
\end{equation}
The names ``triplet'' and ``singlet'' are used by analogy with
the $q\bar q$ isospin state $|u\bar d\rangle$, $|d\bar u\rangle$,
$|u\bar u\rangle \mp |d\bar d\rangle$.
The t-channel analysis according to the
(\ref{triplet-T}-\ref{singlet}) basis
was introduced by Mekhfi \cite{mekhfi}, and led us to re-discover
the quark transverse spin \cite{A+M}.
A similar t-channel analysis applies to
color correlations among the partons \cite{t-color}.

\begin{figure}
\centering\epsfig{file=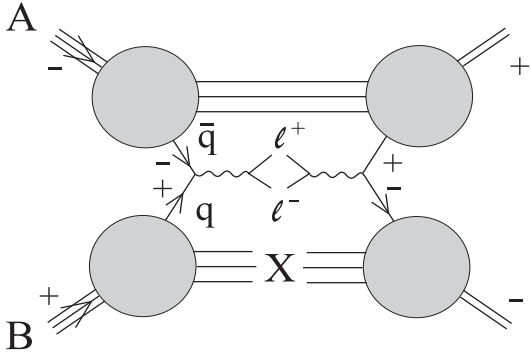,width=7cm,height=5cm}
\caption{Diagram of the Drell-Yan reaction (\ref{D-Y}).}
\end{figure}

\newpage
\noindent
{\bf 8. Selection rules}

\noindent
Due to chiral conservation in hard subprocesses 
(up to corrections of order $m_q/Q$), 
helicity is conserved along the quark lines of the unitarity
diagram. As a first result, the transverse spin asymmetry
in fully inclusive DILS vanishes. In the hand-bag diagram of Fig.2,
it is indeed impossible to have helicity flip
and helicity conservation at the same time.
On the other hand, a double transverse spin asymmetry involving {\it two}
baryons is possible, for instance  
in Drell-Yan reaction (Fig.3) and in semi-inclusive DILS like
\begin{equation}
e^- + \uparrow\! p \to e^- + \uparrow\! \Lambda+X \,,
\label{lambda}
\end{equation}
with $\Lambda$ in the current jet \cite{Baldra,Dourdan,conf,Jaf-lamb}
(this reaction is obtained by {\it crossing} from reaction
(\ref{D-Y}) in which $B = \bar\Lambda$).
Comparing the quark helicity flows in the unitarity diagrams
of Figs. 2 and 3, we can draw general selection rules for probing 
quark transversity in polarized baryons, at leading twist~:

\begin{itemize}
\item{} {\it one needs at least two (initial or final) 
transversely polarized baryons}. One hadron and one lepton does not work. 
Contrarily to the helicity case, no transversity information 
can be exchanged via vector bosons between different fermions lines.
This is the reason why $\delta q(x)$ has not been measured up to now.
\item{} {\it in the unitarity diagram, each t-channel 
baryon-antibaryon pair} (representing {\it one} transversely polarized baryon) 
{\it has to be connected to one (or more) similar pair(s) by two quark lines}.
These lines carry the spin information.
\item{} {\it for fixed spin orientations, 
one must not integrate over the azimuth of the scattering plane} 
(Hikasa theorem \cite{Hikasa}). 
\end{itemize}

\noindent
Other examples of reactions fulfilling these rules have been proposed in
\cite{A+M,conf,ji-probes}.
As final polarized baryons, the choice is pratical restricted 
to hyperons ($\Lambda, \Sigma...$), which are self-analyzing. 
We shall see however in \S 10 that they can be replaced by mesons,
looking at an asymmetry in their azimuthal distributions. 
By-passing the above rules is also possible at non-leading twist \cite{Ja-Ji-e}.

\begin{figure}
\centering\epsfig{file=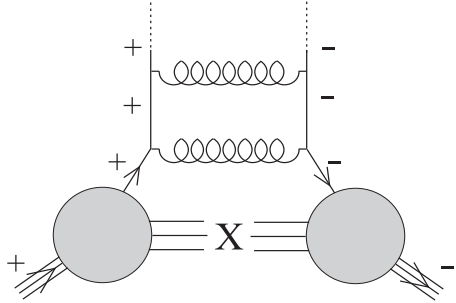,width=6cm,height=4cm}
\caption{Evolution of $\delta q(x,Q^2)$.}
\end{figure}

\newpage
\noindent
{\bf 9. Evolution with $Q^2$}

\noindent
Like $q(x)$ and $\Delta q(x)$, $\delta q(x)$ depends on the 
resolution length $1/Q$ of the subprocess. In the leading
logarithm approximation, the evolution results from the ladder
diagram of Fig.4, where the left half represents the amplitude of
successive gluon emissions and the right half, the complex conjugate
amplitude. For the reasons explained in \S 7, 
the helicities in the left and right parts are opposite.
The evolution of the n$^{th}$ moment of $x^{-1} \; \delta q(x)$ 
is given by 
$$
\delta q^{(n)} (Q^2) = \delta q^{(n)} (Q_0^2) \, \exp\left[
\delta P_n \int_{Q_0^2}^{Q^2} {\alpha_s(Q'^2) \over Q'^2} \, dQ'^2 \right] \,,
$$
with, in the leading logarithm approximation \cite{A+M},
$$
\delta P_n={4\over3}\left({3\over2}-2\sum^n_{j=1}\right)
\,.
$$

\begin{figure}
\centering\epsfig{file=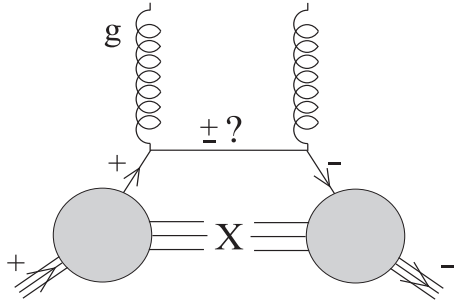,width=6cm,height=4cm}
\caption{Diagram showing the impossibility to couple
$\delta q(x)$ to gluons.}
\end{figure}

The selection rule of \S 8 forbids any coupling to gluon 
distributions (Fig.5), like for the non-singlet
parts of $q(x)$ and $\Delta q(x)$. 
Coupling $\delta q(x)$ to $\delta \bar q(x)$ occurs
at next-to-leading order \cite{NLO} (see Fig.6).

All the $\delta P_n$ for $n \ge 1$ are negative, i.e. all the moments,
in particular the tensor charge $\delta \charge = q^{(1)} - \bar q^{(1)}$, 
are decreasing functions of $Q^2$.
This is unlike the non-singlet parts of $\charge$ and $\Delta \charge$, 
which are conserved.

\newpage

To be independent of the resolution scale $1/Q$, Eq.(\ref{edm}) implies that
$\delta\charge$ and the e.d.m. of the quark evolve in opposite ways.
This can be checked, for instance, from formula (22) of \cite{gavela}.
The quark transversity is "diluted" at high $Q^2$
by gluon radiation, the quark e.d.m. is "screened" at low $Q^2$ by dressing 
with gluons, in such a way that their product is invariant, as it should be
for a physical quantity.

\begin{figure}
\centering\epsfig{file=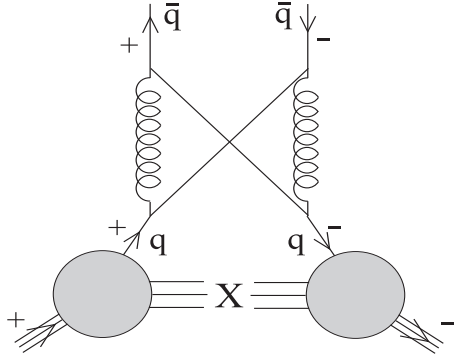,width=6cm,height=5cm}
\caption{Mixing of $\delta q(x)$ with $\delta \bar q(x)$.}
\end{figure}

\medskip
\noindent
{\bf 10. Quark polarimetry}

\noindent
Transversity is transmitted by the quark line in reaction (\ref{lambda})
as follows~:
$$
\ST(q)
=[\delta q(x)/q(x)]\, \ST({\rm proton})
\,,
$$
\begin{equation}
\ST(q') = \hat D_{NN}(\hat\theta) \ {\cal R}\,\ST(q)
\,,
\label{depol}
\end{equation}
$$
\ST(\Lambda)=
[\delta f_{i\to\Lambda}(z) / f_{i\to\Lambda}(z)] \ \ST(q')
\,.
$$
$\hat D_{NN}(\hat \theta) = -2\,\hat s \hat u / (\hat s^2 + \hat u^2)$
is the {\it depolarization parameter} of $q+e^-$ scattering~;
${\cal R}$ is the rotation about the normal to the scattering plane which
transforms the initial quark momentum into the final one~;
$\delta f_{i\to\Lambda}(z)$ is the transversely polarized framentation function
of the quark into a $\Lambda$.
Note that (\ref{depol}) is invariant under cardan transformation.

If the goal of (\ref{lambda}) is just to measure $\delta q(x)$, 
then $\delta f_{i\to\Lambda}(z)$ is only considered as an analyzer of 
the final quark polarization. On the other hand, $\delta f_{i\to\Lambda}(z)$ 
is interesting by itself. From SU(6), one naively expects that it is large for
$s$-quark and small for $u$- and $d$- quarks. But SU(6) is non-relativistic
and quark fragmentation is essentially relativistic...
Anyway, $\delta f_{i\to\Lambda}(z)$ has to be measured 
independently in \cite{CGJJ,conf}
\begin{equation}
e^+\,e^-\to q + \bar q \to \;\uparrow\! \Lambda+\uparrow\! \bar\Lambda 
+ {\rm anything,}
\end{equation}
with $\Lambda$ and $\bar\Lambda$ in opposite jets.
The quark and antiquark 
transversities are correlated as follows 
\begin{equation}
< (\ST^q \cdot \nv)\, (\ST^{\bar q} \cdot \nv') >
\, = \, \hat A_{NN}(\theta) \ \nv' \cdot {\cal R} \nv
\label{Bell}
\end{equation}
\noindent
for any pair of transversity axes $\nv$ and $\nv'$.
${\cal R}$ is the rotation about the normal to the
scattering plane which brings $\pv^q$ along $\pv^{\bar q}$
(cardan invariance is satisfied). Then,

\newpage

$$
< (\ST^\Lambda \cdot \nv)\, (\ST^{\bar\Lambda} \cdot \nv') >
\, = {\sum_i \delta f_{i \to \Lambda} (z) 
\ \delta f_{\bar \imath \to \bar\Lambda} (z') 
\over 
\sum_i f_{i \to \Lambda} (z) 
\ f_{\bar \imath \to \bar\Lambda} (z') }
\ \hat A_{NN}(\theta)  \ \nv' \cdot {\cal R} \nv 
$$
($i=u,d,s...$). Far below the $Z^0$, where the photon exchange dominates,
$\hat A_{NN}$ is given by (\ref{Ann}).
Note that when $|\hat A_{NN}| \simeq 1$, (\ref{Bell}) violates 
the Bell inequality.
On the $Z^0$ pole, $\hat A_{NN}$ is (\ref{Ann}) times 
$-0.74$ for $u$ and $c$ quarks or 
$-0.35$ for $s,d$ and $b$ quarks.

The statistical efficiency of $\Lambda$ (or any other weak-decaying hyperon)
as a quark polarimeter is poor, due to its relatively small abundance 
in quark jets. It is possible however to make a quark polarimeter with mesons.
First, Nachtmann \cite{nacht} and Efremov \cite{hand} 
proposed to guess the quark {\it helicity} from
the sign of $\pv^{(1)} \cdot (\pv^{(2)} \times \pv^{(3)})$,
where $\pv^{(1)}$, $\pv^{(2)}$ and $\pv^{(3)}$ 
are the three leading meson momenta.
Then, Collins \cite{Collins} showed that, for {\it transversity}, 
two mesons (or one meson and the quark momentum) are sufficient.
As an example, a quark moving ``horizontally'' (along $+\zu$) and  polarized 
``upwards'' (along $+\yu$) would emit the leading meson preferentially 
on the ``left'' side of the jet ($p_x >0$) (see also \cite{Baldra}, \S3.7).
This is reminiscent of the Magnus effect in ping-pong or tennis.
Assuming {\it local compensation of transverse momentum} \cite{LCPT},
the sub-leading meson would prefer the opposite side ($p_x < 0$), and so on.
In rapidity $\otimes$ transverse momentum space, 
we obtain a {\it sheared jet}, like 

$$
\uparrow^{p_x}{
\quad\ \, \bullet\quad\quad\ \bullet\quad\quad\ \bullet\quad\quad
\ \bullet\ \quad\quad\,\bullet\quad\quad\ \bullet\quad\quad\ \ \bullet\quad\quad
\over
\bullet\quad\quad\bullet\quad\quad\bullet\quad\quad
\bullet\quad\quad\,\bullet\quad\quad\ \bullet\quad\quad\;\ \bullet\quad\quad}
\!\!\!\!> \log(p_z)
$$
\medskip

\noindent
The {\it sheared} one-particle fragmentation function
can be written in the form~:
\begin{equation}
dN(\uparrow\! q\to h + \X) =
dz \, d^2 \pv\T \, D(z,p_T) 
\left[\, 1 + A_C (z,p_T) 
\, {\ku \times \pv\T \over p_T}
\cdot \ST^q
\right]
\,,
\label{Collins-1}
\end{equation}
where $\kv$ is the quark momentum, $\ku = \kv/k$ ~and
\begin{equation}
z = p/k \,,\quad \pv\T \simeq \pv - p \; \ku
\,.
\label{kine}
\end{equation} 
$A_C(z,p_T) \in [-1,+1]$ is the analyzing power.%
\footnote{
the product $D\;A_C$ is sometimes noted 
$m^{-1} p_T \, H_1^\perp(z,p_T)$.
}
It vanishes when $z \to 0$, because higher rank mesons have less and less 
memory of the leading quark spin.
The sheared-jet (or Collins) effect originates in the non-perturbative 
hadronization of the quark. At high $Q^2$, the latter is preceded 
by a perturbative cascade of gluon emissions. 
So, in (\ref{kine}), the ideal $\kv$ is the quark momentum 
at the end of the cascade.
However, this $\kv$ is difficult to know.
Usually, one replaces it by the "jet momentum"
$\kv^{(jet)}$, which is the quark momentum at the begining of the QCD 
cascade, for instance in DILS $\kv^{(jet)} = \pv^e - \pv'^e$. 
This leads to an underestimation of $z$ with respect to the ideal one.
More deplorable, each emitted gluon changes the quark direction,
introducing a random error on $\pv\T$. 
At high $Q^2$ the one-particle Collins effect becomes blurred 
(see D. Boer, p.258 of \cite{Zeuthen}).
One can avoid this blurring by considering the 
relative Collins effect between {\it two} fast particles of the jet 
\cite{CHL,ji,calib2,jaf-inter},

\newpage

\begin{equation}
dN^{(q\to h_1h_2+\X)} =
dZ \, d\xi \, d^2 \rv\T \, D(Z,\xi,r_T) \,
\left[\, 1 + A_C (Z,\xi,r_T) 
\, {\ku \times \rv\T \over r_T}
\cdot \ST^q
\right]
\,,
\label{Collins-2}
\end{equation}
where $Z = z_1+z_2$, ~$\xi = (z_1 - z_2) /Z$ ~and
$$
\rv\T = {z_2{\pv_1}\T - z_1{\pv_2}\T \over z_1+z_2}
\,.
$$
The above expression for the relative transverse momentum $\rv\T$ of 
$h_1$ and $h_2$ is practically not affected by errors on $\ku$.
Integration is made over the total transverse momentum 
$\Pv\T = {\pv_1}\T + {\pv_2}\T$ of the pair. 
There is also a global Collins effect in $\Pv\T$, but blurred at high $Q^2$.

Note that, apart from the above-mentioned dilution,
one- and two-particle Collins effects are linked.
Due to local compensation of transverse momentum,
the one-particle Collins effect generates a two-particle effect, and vice-versa.
Prescriptions for the calibration of the one and two-particle Collins effects
are given in \cite{calib1} and \cite{calib2}.

Sheared-jet effect occurs in a semi-classically model
of string fragmentation, if we assume that the $q\bar q$ pairs
are created in the $^3\!P_0$ state \cite{A+C+Y} but not 
by the Schwinger mechanism 
\cite{A+C}.
The same model predicts opposite Collins effects for the pion
and the $\rho$ meson \cite{cz}

In S-matrix theory, Collins effect 
results from the interference between a helicity-flip amplitude 
and a helicity-conserving one having different phases,
whence the name of {\it interference fragmentation} \cite{jaf-inter}
also given to the product $DA_c$ in (\ref{Collins-2}).
The non-zero phases may be provided by one or several resonances in the 
two-particle final state. The interference may take place between 
a resonance and a background amplitude \cite{Baldra,CH-unp},
between two different resonances like $\rho$ and $\sigma$ \cite{jaf-inter} 
or between the longitudinal and a transverse state of the same vector meson 
\cite{ji}.

The later case is noteworthy~: consider, for instance, the fragmentation
$q \to \rho^0+\X \to \pi^+\pi^- +\X$ with $\ST^q = \yu$. 
Here the two-particle Collins effect cannot come from the 
spin of the $\rho$ (= "vector" polarization), but from an 
{\it oblique linear} polarization 
($ = T^{zx}$ component of the "tensor" polarization).
The decay angular distribution would be in $1 + a \, (\hat\rv \cdot \uu)^2$,
where $\hat\rv$ is the $\pi^+$ direction in the $\rho^0$ frame
and $\uu = \zu\cos\alpha+\xu\sin\alpha$ is an oblique vector.
Then $A_C (Z,\xi,r_T)$ is odd and $D(Z,\xi,r_T)$ is even in $\xi$ 
(one should {\it not} average over $\xi$ !).
Diagrams giving this effect are shown in Fig.7. There, the interference is
between the $\gamma^\mu$ and $\sigma^{\mu\nu}$ quark-meson form factors 
(equivalently, between the $^3\!S_1$ and $^3\!D_1$ states 
in the $q\bar q$ channel).

\begin{figure}
\centering\epsfig{file=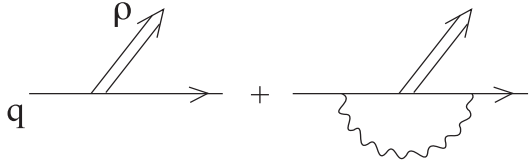,width=7cm,height=2.5cm}
\caption{Diagrams leading to an oblique polarization of the 
$\rho$ meson.}
\end{figure}

\newpage
\noindent
{\bf 11. Single-spin asymmetries in inclusive experiments} 

\noindent
The inclusive experiments with a polarized proton beam \cite{adams}
\begin{equation}
\uparrow\! p + p \longrightarrow  \pi + {\rm anything}
\label{E704}
\end{equation}
have indicated that, for $\pv\T^{(pion)}$ in the $\xu$ direction,
the cross section is larger when the proton spin is along
$+\yu$ rather than $-\yu$.
Assuming that the pion comes from a polarized valence quark 
of the projectile, Collins effect offers a natural explanation 
\cite{Collins,A+C+Y}.
A rough estimation of the asymmetry is 
\begin{equation}
A_N \sim {\delta q(\bar x)\over q(\bar x)} \, 
A_C (\bar z,|\pv\T - \bar z \; \bar \kv'\T|)
\,,
\label{ACY}
\end{equation}
where $x=k_z/p^{(beam)}$, ~$z=p_z^{(pion)}/k'_z$ and 
\begin{equation}
\kv'\T = \kv\T + \Qv\T
\label{bias}
\end{equation} 
is the scattered quark transverse momentum 
and the {\it bar} above $x$, $z$ and $\kv'\T$ indicates the most probable 
values of these quantities, which belong to the internal quark line.
Due to the trigger bias, $\pv\T - \bar z \; \bar\kv'\T$ points toward $\pv\T$.

Another model \cite{Anselmino} 
involves the {\it Sivers effect} \cite{Sivers}, 
where the intrinsic quark momentum $\kv\T$ would be preferably in the direction 
of $\ST^{(proton)} \times \zu$.
This effect is analogous to the Collins one. It was first discarded by
Collins on the basis of time reversal invariance \cite{Collins}.
However Brodsky, Hwang and Schmidt \cite{Brodsky} 
showed than an effective Sivers asymmetry occurs at leading twist in 
semi-inclusive DILS, due to a final state interaction between the scattered 
quark and the spectators.
Collins himself \cite{Collins-dedit} (see also \cite{Ji-Yuan}) 
revisited the consequences of time reversal and agreed with the 
authors of \cite{Brodsky}.
Whether or not we can call it "Sivers effect" is a matter of taste~:
in the Drell-Yan process, there is an analogous effect,
due to an initial state interaction, but with an opposite sign. 
Therefore this mechanism of asymmetry cannot be described in terms of a 
universal (i.e., independent of the subprocess) parton density
[I am indebted to S. Troshin who kindly pointed me out this new development 
about parton distributions]. 

Let us also mention a model combining the relativistic quark orbital motion 
(cf. end of \S2) with an assumed hadron opacity \cite{Boros}, or with
a string traction effect \cite{Burkardt}.

If one relies on the Collins effect only, 
the Soffer inequality makes it rather hard for Eq.(\ref{ACY})
to reproduce the large observed asymmetry \cite{reass}.
However, the introduction of the $\kv\T$ - dependent polarized quark density
${h_1}\T^\perp(x,k_T)$ can considerably improve the situation (see \S12).

\smallskip
Another type of single-spin asymmetry is the spontaneous polarization
of inclusively produced $\Lambda$'s in the direction of $\pv\T \times \zu$
\cite{Felix}.
Many theoretical explanations have been proposed 
(see \cite{Troshin} and ref.4 of \cite{Felix}).
Assuming that the $\Lambda$ comes from the fragmentation of a quark,
its polarization may be viewed as the crossed version of the Sivers effect,
the {\it polarizing fragmentation function} \cite{ABAM}.
Due to the presence of complex amplitudes, it is not forbidden by T-invariance.

\newpage
\noindent
{\bf 12. Quark distributions dependent on both spin and $\kv\T$}

\noindent
At fixed $x$ and $\kv\T$, one can obtain the quark {\it density matrix} 
in spin space from Eqs.(\ref{distrib} - \ref{matriX})~:
\begin{equation}
\rho^q \equiv {1 + \Sv^q \cdot \vec\sigma \over 2}
= { \sum_\X \X \, \rho^B \, \X^\dagger 
\over \sum_\X  {\rm tr} \, \{\X \, \X^\dagger\} }
\label{trace}
\,.
\end{equation}
Let us again consider the (quark + scalar diquark) model of the baryon.
From (\ref{qq-vertex}) and (\ref{trace}), 
after some $\sigma$-matrix algebra, one gets

\begin{equation}
\SL\q = { (1-v^2) \, \SL\uB + 2 \; \vv \cdot \ST\uB \over 1+v^2 }
\,,
\label{SL}
\end{equation}
\begin{equation}
\ST\q = {
\ST\uB + (\wv \cdot \ST\uB) \, \wv 
- (\vv \cdot \ST\uB) \, \vv - 2\SL\uB \, \vv
\over 1+v^2 }
\,,
\label{ST}
\end{equation}
with $\vv = \kv\T / (m_q + x m_B)$ and $\wv = \zu \times \vv$.
Due to the unique state $|\X\rangle$, no spin information
is hidden in the spectator, hence $\rho^q$ has the same purity
as $\rho^B$, that is to say  $|\Sv^q| = |\Sv^B|$.
When one integrates over $\kv\T$, the first terms in the numerators of 
(\ref{SL}) and (\ref{ST}) give the already known $\Delta q(x)$ and 
$\delta q(x)$, and the other terms disappear. However, at fixed $\kv\T$,
three new polarized quark distributions appear~:
\begin{itemize}
\item{} $g_{1T} (x,k_T) = q(x,k_T) \times $ second term of (\ref{SL}),
\item{} $h_{1T}^\perp (x,k_T) = q(x,k_T) \times $ 
second + third terms of (\ref{ST}),
\item{} $h_{1L}^\perp (x,k_T) = q(x,k_T) \times $ last term of (\ref{ST}).
\end{itemize}

\noindent
These new objects where introduced by Tangerman and Mulders \cite{tang-mul},
and independently by Kotzinian \cite{kotz}.
They obey positivity bounds analogous to the Soffer inequality \cite{BBHM}, 
obtained from the positivity of the quark-baryon correlation 
matrix (\ref{correl}).

Taking $h_{1T}^\perp$ into account can considerably improves  the
explanation of the single-spin asymmetry in (\ref{E704}) by the Collins effect,
for the following reasons 
(X. Artru, Refs.\cite{RIKEN} p.325, \cite{Zeuthen} p.300)~:
1) $\kv\T$ of (\ref{bias}) most probably points toward $\pv\T$ (trigger bias)~;
2) for $\kv\T \perp \ST\uB$ the first two terms of (\ref{ST}) 
reinforce each other, giving $\ST\q = \ST\uB$, 
instead of $\ST\q = \ST\uB /(1+v^2)$ for the first term alone.
This by-passes the Soffer bound, but it should be remembered
that the latter is valid only after $\kv\T$ integration.

\medskip
\noindent
{\bf 13. First signature of quark transversity}

\noindent
Until recently, the only indications of a nonvanishing $\delta q(x)$
were the single-spin asymmetries like (\ref{E704}).
However, due to the not very high $\pv\T$, 
these asymmetries could also be explained 
by pure hadronic models \cite{sof-torn,barni} 
not involving quarks explicitely.

Early proposals \cite{kunne,HELP} 
were made to measure $\delta q(x)$ in the semi-inclusive DILS 
reactions like (\ref{lambda}) or 
\begin{equation}
e^\pm \, + \uparrow\! p \to e^\pm + {\rm meson} + {\rm anything}
\,.
\label{semi}
\end{equation}
Reaction (\ref{lambda}) uses the $\Lambda$ and (\ref{semi}) the Collins effect 
as quark polarimeters. These proposals were not realized,
but the principles were taken up for HERMES Run 2 
(M.G. Vincter, p.311 of \cite{Zeuthen}) 
and run 2003 of COMPASS (E.-M. Kabu\ss, p.323 of \cite{Zeuthen}).
By the way, participating to these proposals with enthusiastic and 
hard-working experimental teams was really stimulating.

Reactions with two polarized beams like (\ref{D-Y}) are planned 
at RHIC collider. However, most of them involve $q + \bar q$ subprocesses,
thus have $\delta\bar q(x)$, which is presumably small, in factor. 
On the other hand, the $u+u \to u+u$ and $d+d \to d+d$ subprocesses
have $\hat A_{NN}$ less than $1/11$ \cite{qq-scat, A+M, Sivers2}.

The first evidence of quark transversity was reported in 2000-2001 
by the HERMES collaboration \cite{hermes}, in reaction (\ref{semi}).
The target was longitudinally, not transversally polarized~;
nevertheless, event by event, the proton spin had some transverse component 
with respect to the virtual photon momentum $\Kv$, which 
makes an angle $\simeq (E'_e/E_e)^{1\over2} 2Mx/Q$ with the electron beam. 
Measuring the pion azimuthal angle $\phi$ around $\Kv$, 
they recorded an azimuthal asymmetry in $\sin\phi$, at the 4\% level.
It could be explained in terms of $\delta q(x)$ times 
the one-particle Collins effect 
\cite{fit-hermes}.
But here contributions from non-leading twist functions like $h_L(x)$
are of the same order in $1/Q$ \cite{Boer-Oganess}.
Furthermore, as in the case of (\ref{E704}), $\delta q(x)$ could be reinforced 
by ${h_1}\T^\perp(x)$ (see end of \S 12). 

\medskip
\noindent
{\bf 14. Conclusions}

\noindent
The aim of this introduction, which is far from exhaustive, 
was to give an aper\c cu of the richness of 
transverse spin and its differences with helicity. 

From the experimental point of view, the physics of quark
transversity in deep inelastic reaction is still practically unexplored.
This situation will certainly change rapidly, with planned experiments
at DESY (HERMES), Brookhaven (RHIC) and CERN (COMPAS), 
but there is a long way before knowing $\delta q(x)$ 
as precisely as $\Delta q(x)$ now. 
Unless polarized anti-proton beams become feasible,
experiments probing quark transversity  will rely mainly 
on ``quark polarimeters'', like $\Lambda$'s or the Collins effect.
These polarimeters will have to be calibrated at $e^+e^-$ colliders.
 
The Collins polarimeter will by the way allow the flavor
decomposition of $\delta q(x)$, using mesons of various charges 
and strangnesses.
Quark polarimetry is by itself an interesting topic of non-perturbative QCD,
and may teach us something about the breaking of chiral symmetry.
Let us recall that, if chiral symmetry were unbroken, 
transversity would be undefined.

The transversity physics program is not at all a ``remake'' of the 
helicity one. Helicity and transversity probe rather different aspects 
of the hadron structure.
Differences between $\Delta q(x)$ and $\delta q(x)$ will
reveal non-relativistic effects in the baryon wave function.
Also $\delta q(x)$ does not couples to gluon distributions, 
thus is free from anomaly. In that respect it is a more clean probe
that $\Delta q(x)$.

In fact, the combination of helicity and transversity measurements
will perhaps be the most interesting.
Polarized parton densities taking only the helicity degree of freedom are 
almost "classical". Quantum aspects of spin correlations, like violation of
Bell's inequality, can be found only when varying the spin 
quantification axis. Classical-like densities in one basis appear as quantum
interferences in the other basis. 

Transversely polarized experiment may also be a tool 
to detect new physics \cite{Hikasa,mekhfi-new}.

On the theoretical side, much work has been done
(see Ref.\cite{barone}), but much remains also to be done. 
Which reggeon, if any, governs $\delta q(x)$ at low $x$ has not been discussed,
to our knowledge. 
The connection between the tensor charge of the baryons and their magnetic 
and electric-dipole moments has to be clarified.


\end{document}